\documentclass[onecollarge]{svjour2}
\smartqed  
\usepackage{graphicx}
\journalname{Few-Body Systems}
\begin{document}

\title{\boldmath Hyperons  analogous to the $\Lambda(1405)$
}


\author{Yongseok Oh
}


\institute{Y. Oh \at
              Department of Physics, Kyungpook National University, Daegu 702-701, Korea \\
              \email{yohphy@knu.ac.kr}           
}

\date{Received: date / Accepted: date}

\maketitle

\begin{abstract}
The low mass of the $\Lambda(1405)$ hyperon with $j^P = 1/2^-$, 
which is higher than the ground state $\Lambda(1116)$ mass
by 290~MeV, is difficult to understand in quark models.
We analyze the hyperon spectrum in the bound state approach of the Skyrme model
that successfully describes both the $\Lambda(1116)$ and the $\Lambda(1405)$.
This model predicts that several hyperon resonances of the same spin but with opposite parity
form parity doublets that have a mass difference of around 300~MeV,
which is indeed realized in the observed hyperon spectrum.
Furthermore, the existence of the $\Xi(1620)$ and the $\Xi(1690)$ of $j^P=1/2^-$
is predicted by this model. 
Comments on the $\Omega$ baryons and heavy quark baryons are made as well.
\end{abstract}

\section{Introduction}

The low mass of the $\Lambda(1405)$ hyperon has been a puzzle
when it is described as a $P$-wave
three-quark state~\cite{IK78a,AMS94}. 
Instead, interpreting the $\Lambda(1405)$ as a $\bar{K}N$ bound state has been
successful to understand various properties of the 
$\Lambda(1405)$~\cite{DT59,AS62,VJTB85}.
It is then natural to search for other hyperons that have similar structure as
the $\Lambda(1405)$.
In this paper, we investigate hyperon spectrum in the bound state approach in the
Skyrme model to search for the hyperons analogous to the $\Lambda(1405)$.
We will also make a short comment on the heavy-flavored analog of the $\Lambda(1405)$.

In the bound state approach to the Skyrme model~\cite{CK85}, hyperons are described as 
bound states of the soliton and mesons carrying strangeness
quantum number such as the kaon and the $K^*$ vector mesons.
The underlying dynamics between the soliton and kaon is described by the Lagrangian of meson
degrees of freedom.
As shown in Ref.~\cite{CK85}, the Wess-Zumino term in an SU(3) chiral Lagrangian
pushes up the $S=+1$ state and pulls
down the $S=-1$ state. 
As a result,  the $S=+1$ pentaquark $\Theta^+$ cannot be a bound state,
while the $S=-1$ states correspond to the normal hyperons. We refer to Ref.~\cite{PRM04} for further
discussions on the $\Theta^+$ in this model.
Furthermore, this model renders two kinds of bound state, one in $P$-wave and
one in $S$-wave. The $P$-wave state is strongly bound and, when quantized, it gives the
ground states of hyperons with $j^P = 1/2^+$ and $3/2^+$.
On the other hand, the $S$-wave state is an excited state and, when quantized, it
corresponds to the $\Lambda(1405)$ with $j^P = 1/2^-$.
Therefore, this model gives a natural way to describe both the $\Lambda(1116,1/2^+)$ 
and the $\Lambda(1405,1/2^-)$ on the same 
ground~\cite{SSG95}.

Here, we present the results on the hyperon spectrum based on this approach.
We found that there exists a pattern in the hyperon spectrum, which can be verified by
the measured hyperon masses. In particular, the predicted $\Xi$ and $\Omega$ spectra 
are very different from the quark model predictions and can explain several
puzzles in hyperon spectrum, which can be further tested by experiments at current accelerator
facilities.

\section{Hyperon Spectrum}

There have been many models on the structure of hyperons, mostly based on quark dynamics.
As shown in Table~\ref{tab:xi-omega}, the hyperon mass spectrum have been predicted by
various models. It reveals that the predictions are highly model-dependent and all these quark models
leave a puzzle on the low mass of the $\Xi(1620)$ and the $\Xi(1690)$, which is very similar to the puzzle of
the $\Lambda(1405)$.
In the bound state approach in the Skyrme model, the mass of a hyperon with isospin $i$ and spin $j$ is written as~\cite{Oh07} 
\begin{eqnarray}
M(i,j,j_m^{}) &=& M_{\rm sol} + n_1^{} \omega_1^{} + n_2^{} \omega_2^{}
\nonumber \\ && \mbox{}
+ \frac{1}{2\mathcal{I}} \Biggl\{
i(i+1) + c_1^{} c_2^{} j_m^{}(j_m^{}+1)
+ (\bar{c}_1^{} - c_1^{} c_2^{}) j_1^{}(j_1^{}+1)
+ (\bar{c}_2^{} - c_1^{} c_2^{}) j_2^{}(j_2^{}+1)
\nonumber \\ && \mbox{} \qquad +
\frac{c_1^{}+c_2^{}}{2} [j(j+1) - j_m^{}(j_m^{}+1) - i(i+1)]
+ \frac{c_1^{} - c_2^{}}{2} \, \vec{R} \cdot (\vec{J}_1 - \vec{J}_2)
\Biggr\},
\label{eq:mass0}
\end{eqnarray}
where $\vec{J}_1$ and $\vec{J}_2$ are the grand spins of the $P$-wave and $S$-wave kaon, respectively,
and $\vec{J}_m = \vec{J}_1 + \vec{J}_2$. The total spin of the system is then given by 
$\vec{J} = \vec{J}_{\rm sol} + \vec{J}_m$, where $\vec{J}_{\rm sol}$ is the soliton spin. 
The number and energy of the bound kaons are $n_i$ and $\omega_i$,
respectively, and $c_i$ are the hyperfine splitting constants of the bound states.
This mass formula consists of three parts. The soliton mass $M_{\rm sol}$ is of $O(N_c)$, where $N_c$
is the number of color, the energy of the bound kaon is of $O(N_c^0)$, and the hyperfine term is of $O(1/N_c)$.
Therefore, the mass splitting between the $\Lambda(1405)$ and the $\Lambda(1116)$ mainly comes from 
the energy difference between the $P$-wave kaon and the $S$-wave kaon, 
In the simple model of Ref.~\cite{CK85},
$\omega_2 - \omega_1$ was estimated to be about 200~MeV~\cite{SSG95},
while its empirical value is about 300~MeV.

\begin{table}[t]
\caption{\label{tab:xi-omega} Low-lying $\Xi$ and $\Omega$ baryon spectrum of spin $1/2$ and
$3/2$ predicted by the non-relativistic quark model of
Chao et al. (CIK)~\cite{CIK81}, relativized quark model of Capstick and Isgur
(CI)~\cite{CI86}, Glozman-Riska model (GR)~\cite{GR96b}, 
large $N_c$ analysis~\cite{CC00,SGS02,GSS03,MS04b,MS06b},
algebraic model (BIL)~\cite{BIL00}, QCD sum rules (QCD-SR)~\cite{LL02},
and the recent nonrelativistic quark model of Pervin and Roberts (PR)~\cite{PR07}.
The mass is given in the unit of MeV.}
\centering
\begin{tabular}{cccccccc} \hline\noalign{\smallskip}
State & CIK  & CI  & GR  &
Large-$N_c$  & BIL  &
QCD-SR  & PR \\[3pt]
\tableheadseprule\noalign{\smallskip}
$\Xi(\frac12^+)$ & $1325$ & $1305$ & $1320$ &        & $1334$ &
$1320$  & $1325$ \\
                 & $1695$ & $1840$ & $1798$ & $1825$ & $1727$ & & $1891$   \\
                 & $1950$ & $2040$ & $1947$ & $1839$ & $1932$ &  & 2014   \\ \hline
$\Xi(\frac32^+)$ & $1530$ & $1505$ & $1516$ &        & $1524$ &  & $1520$  \\
                 & $1930$ & $2045$ & $1886$ & $1854$ & $1878$ &   & $1934$ \\
                 & $1965$ & $2065$ & $1947$ & $1859$ & $1979$ &  & $2020$  \\ \hline
$\Xi(\frac12^-)$ & $1785$ & $1755$ & $1758$ & $1780$ & $1869$ &
$1550$ & $1725$ \\
                 & $1890$ & $1810$ & $1849$ & $1922$ & $1932$ & & $1811$   \\
                 & $1925$ & $1835$ & $1889$ & $1927$ & $2076$ &    \\ \hline
$\Xi(\frac32^-)$ & $1800$ & $1785$ & $1758$ & $1815$ & $1828$ &
$1840$ & $1759$ \\
                 & $1910$ & $1880$ & $1849$ & $1973$ & $1869$ &  & $1826$  \\
                 & $1970$ & $1895$ & $1889$ & $1980$ & $1932$ &    \\
\hline
$\Omega(\frac12^+)$ & $2190$ & $2220$ & $2068$ & $2408$ & $2085$ &  & $2175$ \\
                    & $2210$ & $2255$ & $2166$ &        & $2219$ & & $2191$ \\ \hline
$\Omega(\frac32^+)$ & $1675$ & $1635$ & $1651$ &        & $1670$ & & $1656$ \\
                    & $2065$ & $2165$ & $2020$ & $1922$ & $1998$ & & $2170$ \\
                    & $2215$ & $2280$ & $2068$ & $2120$ & $2219$ & & $2182$ \\ \hline
$\Omega(\frac12^-)$ & $2020$ & $1950$ & $1991$ & $2061$ & $1989$ & & $1923$ \\ \hline
$\Omega(\frac32^-)$ & $2020$ & $2000$ & $1991$ & $2100$ & $1989$ & & $1953$ \\
\hline
\end{tabular}
\end{table}

In principle, the mass parameters in Eq.~(\ref{eq:mass0}) should be calculated for a given dynamics
of the meson-soliton system. However, this is highly nontrivial because of the complexity of the
hadron dynamics. Instead, we fit the parameters to some known hyperon masses and predict the masses
of other hyperons. The results obtained in this way are given in Table~\ref{tab:mass}.
In this model, the parity of a hyperon changes if the $P$-wave kaon is replaced by the $S$-wave kaon.
Since the energy difference between the two kaons is about 300~MeV, there are pairs of hyperons of having
same spin and the opposite parity having a mass difference of about 300~MeV.
Since the mass of the ground state of the $\Xi(1/2^+)$ is 1318~MeV, we can expect to have a $\Xi(1/2^-)$ 
state at a mass of about 1618~MeV. 
In our model, there exist two $\Xi$ states of this mass.
This is because the two kaons, one in $P$-wave and one in $S$-wave, can make either $j_m = 0$
or $j_m = 1$. When combined with the soliton spin $j_{\rm sol} = 1/2$, these states give two $j=1/2$ states 
and one $j=3/2$ state.
This explains naturally the existence of two $\Xi$ baryons with $j^P = 1/2^-$ that have similar masses.
In fact, there are candidates for these two $\Xi$ baryons in PDG~\cite{PDG10}: the one-star rated $\Xi(1620)$ 
and the three-star rated $\Xi(1690)$. 
However, since the observation of the $\Xi(1620)$ at early 1980s~\cite{HACN81},
there is no other experimental confirmation of this state. 
Instead, several experiments reported no signal of this state~\cite{PDG10}.
Therefore, it is strongly required to resolve this issue urgently at current experimental facilities.
(See Ref.~\cite{CLAS07b} for a recent experiment for $\Xi$ baryons.)

\begin{table}[t]
\caption{\label{tab:mass}
Mass spectrum of our model.
The underlined values are used to determine the mass parameters.
The values with the $*$ symbol are obtained by considering
the mixing effect. The question mark after the particle name means that
the spin-parity quantum numbers are not identified by the Particle Data Group (PDG).}
\centering
\begin{tabular}{ccc} \hline\noalign{\smallskip}
Particle Name & Mass (MeV) & Assigned State
\\[3pt] \tableheadseprule\noalign{\smallskip}
$N$ & $\underline{939}$ \\
$\Delta$ & $\underline{1232}$ \\ \hline
$\Lambda (\frac12^+)$ & $\underline{1116}$ & $\Lambda(1116)$ \\
$\Lambda (\frac12^-)$ & $\underline{1405}$ & $\Lambda(1405)$ \\
$\Sigma (\frac12^+)$ & $1164$ & $\Sigma(1193)$ \\
$\Sigma (\frac32^+)$ & $\underline{1385}$ & $\Sigma(1385)$ \\
$\Sigma (\frac12^-)$ & $1475$ & $\Sigma(1480)?$ \\
$\Sigma (\frac32^-)$ & $1663$ & $\Sigma(1670)$ \\ \hline
$\Xi (\frac12^+)$ & $\underline{1318}$ & $\Xi(1318)$ \\
$\Xi (\frac32^+)$ & $1539$ & $\Xi(1530)$ \\
$\Xi (\frac12^-)$ & $1616(1614*)$ & $\Xi(1620)?$ \\
$\Xi (\frac12^-)$ & $1658(1660*)$ & $\Xi(1690)?$ \\
$\Xi (\frac32^-)$ & $\underline{1820}$ & $\Xi(1820)$ \\
$\Xi (\frac12^+)$ & $1932$ & $\Xi(1950)?$ \\
$\Xi (\frac32^+)$ & $\underline{2120}$ & $\Xi(2120)?$ \\ \hline
$\Omega (\frac32^+)$ & $1694$ & $\Omega(1672)$ \\
$\Omega (\frac12^-)$ & $1837$ \\
$\Omega (\frac32^-)$ & $1978$ \\
$\Omega (\frac12^+)$ & $2140$ \\
$\Omega (\frac32^+)$ & $2282$ & $\Omega(2250)?$ \\
$\Omega (\frac32^-)$ & $2604$
\\ \hline
\end{tabular}
\end{table}

The above analysis reveals that the $\Xi(1620)$ and the $\Xi(1690)$ are analogue states of the $\Lambda(1405)$.
Recently, the BABAR Collaboration claimed that the spin-parity of the $\Xi(1690)$ is 
$1/2^-$~\cite{BABAR08}, which supports our prediction.
On the other hand, by replacing two $P$-wave kaons in the $\Xi(1382)$ and in the $\Xi(1530)$, we 
predict that the $\Xi(1950)$ has $j^P=1/2^+$ and the $\Xi(2120)$ has $j^P=3/2^+$.
Their spin-parity quantum numbers are not known yet and should be identified by future experiments.

Comparing the predictions presented in Tables~\ref{tab:xi-omega} and \ref{tab:mass} shows that
our prediction on the $\Omega$ hyperon spectrum is drastically different from the quark model
predictions. In quark models, the second lowest $\Omega$ hyperon has a mass of around 2~GeV.
In our model, the second state has a mass of around 1840~MeV and $j^P = 1/2^-$.
Again, we can find that this low mass of the $\Omega$ excited state can hardly be explained by
quark models. Thus, it is very interesting to see whether such low mass $\Omega$ hyperon
really exists.
Furthermore, most quark models predict that the lowest $\Omega$ baryon 
with $j^P=1/2^-$ is degenerate or almost degenerate in mass with the lowest  
$\Omega$ baryon with $j^P = 3/2^-$, which is in contradiction to our predictions.
These inconsistency with quark model predictions can be tested by future experiments.

If we extend our model to heavy quark baryons~\cite{RRS92}, we can also find a similar pattern in charm
and bottom baryon spectra. Here, one should take into account the center-of-mass problem 
because of the heavy mass of the charm or bottom meson. In Ref.~\cite{OP97}, the binding energies of
the soliton--heavy-meson system were calculated in the rest frame of the heavy meson, which shows
that the energy difference between the positive parity state and the negative parity state is again close
to 300~MeV, which can explain the observed mass difference between the $\Lambda_c(2286)$ of $j^P = 1/2^+$
and the $\Lambda_c(2595)$ of $j^P = 1/2^-$. In quark models, the mass difference between the two states
are estimated to be $250 \sim 350$~MeV depending on the details of the model on the quark 
dynamics~\cite{CI86,RP08}. Therefore, more detailed studies are
needed to clarify the structure of the $\Lambda_c(2595)$.

\section{Summary}

We have analyzed hyperon excited states in the bound state approach in the Skyrme model.
This model can explain both the $\Lambda(1116)$ and the $\Lambda(1405)$ on the same footing.
We found that the $\Xi(1620)$ and the $\Xi(1690)$ can be regarded as the analogous states of
the $\Lambda(1405)$. This model also gives predictions on $\Omega$ hyperons that are very
different from quark model predictions. However, there is almost no experimental information on the
spectrum of $\Omega$ baryons. Therefore, detailed studies on the excited states of $\Xi$ and $\Omega$ baryons
at current experimental facilities are highly required.

\begin{acknowledgements}
The author is grateful to B.-Y. Park for fruitful discussions.
This work was supported by
Basic Science Research Program through the National
Research Foundation of Korea (NRF) funded by the Ministry of Education,
Science and Technology (Grant \mbox{No.} 2010-0009381).
\end{acknowledgements}


\begin{thebibliography}{10}

\bibitem{IK78a}
Isgur, N., Karl, G.: $P$--wave baryons in the quark model.
\newblock Phys. Rev. D \textbf{18}, 4187 (1978).

\bibitem{AMS94}
Arima, M., Matsui, S., Shimizu, K.: $\Lambda(1405)$ and meson-baryon interactions in a quark model.
\newblock Phys. Rev. C \textbf{49}, 2831 (1994).

\bibitem{DT59}
Dalitz, R.~H., Tuan, S.~F.: Possible resonant state in pion-hyperon scattering.
\newblock Phys. Rev. Lett. \textbf{2}, 425 (1959).

\bibitem{AS62}
Arnold,  R.~C., Sakurai, J.~J.: Vector mesons and the $KN$, $\bar{K}N$ interactions.
\newblock Phys. Rev. \textbf{128}, 2808 (1962).

\bibitem{VJTB85}
Veit, E.~A., Jennings, B.~K., Thomas, A.~W., Barrett, R.~C.: $S$-wave meson-nucleon scattering in an SU(3) cloudy bag model.
\newblock Phys. Rev. D \textbf{31}, 1033 (1985).

\bibitem{CK85}
Callan, C.~G., Klebanov, I.: Bound-state approach to strangeness in the Skyrme model.
\newblock Nucl. Phys. B \textbf{262}, 365 (1985).

\bibitem{PRM04}
Park, B.-Y., Rho, M., Min, D.-P.: Kaon-soliton bound state approach to pentaquark states.
\newblock Phys. Rev. D \textbf{70}, 114026 (2004).

\bibitem{SSG95}
Schat, C.~L., Scoccola, N.~N., Gobbi, C.: $\Lambda(1405)$ in the bound-state soliton model.
\newblock Nucl. Phys. A \textbf{585}, 627 (1995).

\bibitem{Oh07}
Oh, Y.: $\Xi$ and $\Omega$ baryons in the Skyrme model.
\newblock Phys. Rev. D \textbf{75}, 074002 (2007).

\bibitem{CIK81}
Chao, K.-T., Isgur, N., Karl, G.: Strangeness $-2$ and $-3$ baryons in a quark model with
chromodynamics.
\newblock Phys. Rev. D \textbf{23}, 155 (1981).

\bibitem{CI86}
Capstick, S., Isgur, N.: Baryons in a relativized quark model with chromodynamics.
\newblock Phys. Rev. D \textbf{34}, 2809 (1986).

\bibitem{GR96b}
Glozman, L.~\mbox{Ya}., Riska, D.~O.: The spectrum of the nucleons and the strange hyperons and chiral
dynamics.
\newblock Phys. Rep. \textbf{268}, 263 (1996).

\bibitem{CC00}
Carlson, C.~E., Carone, C.~D.: Predictions for decays of radially excited baryons.
\newblock Phys. Lett. B \textbf{484}, 260 (2000).

\bibitem{SGS02}
Schat, C.~L., Goity, J.~L., Scoccola, N.~N.: Masses of the $70^-$ baryons in large $N_c$ QCD.
\newblock Phys. Rev. Lett. \textbf{88}, 102002 (2002).

\bibitem{GSS03}
Goity, J.~L. , Schat, C., Scoccola, N.~N.: Analysis of the $[56,2^+]$ baryon masses in the $1/N_c$
expansion.
\newblock Phys. Lett. B \textbf{564}, 83 (2003).

\bibitem{MS04b}
Matagne, N., Stancu, \mbox{Fl}.: The $[56,4^+]$ baryons in the $1/N_c$ expansion.
\newblock Phys. Rev. D \textbf{71}, 014010 (2005).

\bibitem{MS06b}
Matagne, N., Stancu, \mbox{Fl}.: Masses of $[70,\ell^+]$ baryons in the $1/N_c$ expansion.
\newblock Phys. Rev. D \textbf{74}, 034014 (2006).

\bibitem{BIL00}
Bijker, R., Iachello, F., Leviatan, A.: Algebraic models of hadron structure.
\newblock Ann. Phys. (N.Y.) \textbf{284}, 89 (2000).

\bibitem{LL02}
Lee F.~X., Liu, X.: Predictive ability of QCD sum rules for excited baryons.
\newblock Phys. Rev. D \textbf{66}, 014014 (2002).

\bibitem{PR07}
Pervin M., Roberts, W.: Strangeness $-2$ and $-3$ baryons in a constituent quark model.
\newblock Phys. Rev. C \textbf{77}, 025202 (2008).

\bibitem{HACN81}
Hassall, J.~K.,  et~al.: Production of $S = -2$ and $-3$ baryon states in $6.5$-GeV/$c$
$K^-p$ interactions.
\newblock Nucl. Phys. B \textbf{189}, 397 (1981).

\bibitem{PDG10}
Nakamura, K., et~al. (Particle Data Group):  Review of particle physics.
\newblock J. Phys. G \textbf{37}, 075021 (2010).

\bibitem{CLAS07b}
Guo, L., et~al. (CLAS Collaboration): Cascade production in the reactions $\gamma p \to K^+K^+(X)$ and
$\gamma p \to K^+K^+\pi^-(X)$.
\newblock Phys. Rev. C \textbf{76}, 025208 (2007).

\bibitem{BABAR08}
Aubert, B., et~al. (BABAR Collaboration): Measurement of the spin of the $\Xi(1530)$ resonance.
\newblock Phys. Rev. D \textbf{78}, 034008 (2008).

\bibitem{RRS92}
Rho, M., Riska, D.~O., Scoccola, N.~N.: The energy levels of the heavy flavour baryons in the
topological soliton model.
\newblock Z. Phys. A \textbf{341}, 343 (1992).

\bibitem{OP97}
Oh, Y., Park, B.-Y.; Solitons bound to heavy mesons.
\newblock Z. Phys. A \textbf{359}, 83 (1997).

\bibitem{RP08}
Roberts, W., Pervin, M.: Heavy baryons in a quark model.
\newblock Int. J. Mod. Phys. A \textbf{23}, 2817 (2008).

\end{thebibliography}

\end{document}